\documentclass[11pt]{article}

\usepackage{PRIMEarxiv}

\usepackage{booktabs}
\usepackage{multirow}
\usepackage{makecell}

\usepackage{times}
\usepackage{courier} % for monospaced text
\usepackage{array}
\usepackage{graphicx}    % needed for including graphics e.g. EPS, PS
\usepackage{epstopdf}
\usepackage{tweaklist}
\usepackage[footnotesize]{caption}
\usepackage[T1]{fontenc}
\usepackage[utf8]{inputenc}
\usepackage{lmodern}
\usepackage{microtype}

\usepackage{booktabs}
\usepackage{tabularx}
\usepackage{longtable}

\usepackage{multirow}

\usepackage{enumitem}
\usepackage{amsmath,amssymb}
\usepackage[round]{natbib}
\setcitestyle{authoryear}
\usepackage{hyperref}
\hypersetup{colorlinks=true, linkcolor=blue, urlcolor=blue, citecolor=blue}

\usepackage{titlesec}
\newcommand{\mainsectionformat}{\titleformat{
\section}{\normalfont\Large\bfseries}{
\thesection}{0.75em}{}}
\newcommand{\appendixsectionformat}{
\titleformat{\section}{\normalfont\Large\bfseries}{
Appendix~\Alph{section}:}{0.75em}{}}
\mainsectionformat

\setlist[itemize]{noitemsep, topsep=2pt}
\setlist[enumerate]{noitemsep, topsep=2pt}

\title{Orchestrating LLM Agents for Scientific Research: A Pilot Study of Multiple Choice Question (MCQ) Generation and Evaluation}
\author{Yuan An\\
College of Computing and Informatics\\
Drexel University \\
Philadelphia, PA 19104, USA \\
\texttt{ya45@drexel.edu}
\date{} % leave blank for journal submission
}

\begin{document}
\maketitle

%%%%%%%%%%%%%%%%%%
%%
%% Abstract
%%
%%%%%%%%%%%%%%%%%%%

\begin{abstract}

Advances in large language models (LLMs) are rapidly transforming scientific work, 
yet empirical evidence on how these systems reshape research activities remains 
limited. We report a mixed-methods pilot evaluation of an AI-orchestrated research 
workflow in which a human researcher coordinated multiple LLM-based agents to 
perform data extraction, corpus construction, artifact generation, 
and artifact evaluation. Using the generation and assessment of multiple-choice 
questions (MCQs) as a testbed, we collected 1,071 SAT Math
MCQs and employed LLM agents to extract questions from PDFs, retrieve and convert 
open textbooks into structured representations, align each MCQ with relevant 
textbook content, generate new MCQs under specified difficulty and cognitive 
levels, and evaluate both original and generated MCQs using a 24-criterion 
quality framework. 
Across all evaluations, average MCQ quality was high (mean 4.64-4.89/5) 
and 74.1-93.4\% of individual criterion scores reached the maximum rating of~5. 
However, criterion-level analysis and equivalence testing show that generated 
MCQs are \emph{not fully comparable} to expert-vetted baseline questions. Strict
similarity (24/24 criteria equivalent) was never achieved, and only 9-12/24 
criteria met an equivalence threshold depending on LLM-based 
judge and analysis track. 
Persistent gaps concentrated in skill\ depth, cognitive engagement, 
difficulty calibration, and metadata alignment, while surface-level qualities, 
such as {grammar fluency}, {clarity options}, {no duplicates}, 
were consistently strong.
Beyond MCQ outcomes, the study documents a labor shift. The researcher's work 
moved from ``authoring items'' toward \emph{specification, orchestration, verification}, 
and \emph{governance}. Formalizing constraints, designing rubrics, building validation loops, 
recovering from tool failures, and auditing provenance constituted the 
primary activities. We discuss implications for the future of scientific work, 
including emerging ``AI research operations'' skills 
required for AI-empowered research pipelines.

\end{abstract}


\thispagestyle{plain}

%%%%%%%%%%%%%%%%%%
%%
%% Introduction
%%
%%%%%%%%%%%%%%%%%%%
\section{Introduction}

Artificial intelligence is moving beyond basic tasks like spell-checking to become a 
core part of the actual research work. LLMs can now handle data, write content, and 
generate code. While many are still debating whether AI output is ``good enough,'' the 
deeper shift is already underway. In some fields, the technology is redefining what 
it means to be a researcher. In these domains, substantial portions of hypothesis 
generation, data collection, analysis, and even validation can be executed by 
computational systems rather than humans. As algorithms take over drafting and 
analysis, the researcher’s role shifts from doing the work to managing AI systems. 
This paper tests that shift. We argue that while AI allows us to produce much more, 
it fundamentally changes scientific labor from creating content to defining and 
verifying it.  This shift is central to the future-of-work conversation in scientific 
occupations. The practical question is no longer whether AI assists research, but 
how it redefines the work itself:
\begin{quote}
When a researcher utilizes an AI-driven pipeline to complete a project end-to-end, 
which tasks are delegated, and what new responsibilities emerge?
\end{quote}

In this study, we attempt to answer this question through a 
research application that is defined as \emph{the large-scale generation and evaluation of SAT Math 
multiple-choice questions (MCQs) derived from open-access textbooks.} Automatically 
generating various artifacts has long been studied as a way for reducing 
human labor demands. In the standardized test preparation industry, AI-driven content generation  
has profound impacts on industry, educators, and learners.

The broader U.S. test preparation sector was projected to grow by more than \$18 
billion between 2024 and 2029 (\cite{Technavio2026USprep}). However, on January 21, 2026, 
Google announced free, full-length digital SAT practice tests within its Gemini model 
family through a partnership with The Princeton Review. Around the same time (January 2026), 
research systems such as ReQUESTA (\cite{tian2026cognitively}) introduce hybrid agentic 
frameworks for generating cognitively diverse multiple-choice questions (MCQs) at scale.

If large language models (LLMs) can produce SAT-style preparation materials that 
are pedagogically and cognitively comparable to those written by human experts, the 
economic and professional structure of the test preparation industry would be  significantly 
affected. This motivates a central empirical question:
\begin{quote}
To what extent are SAT questions generated by LLMs comparable to 
human-authored questions along pedagogical quality, cognitive demand, and assessment 
validity dimensions?
\end{quote}

\begin{figure}[htbp]
\centering
\includegraphics[width=\linewidth]{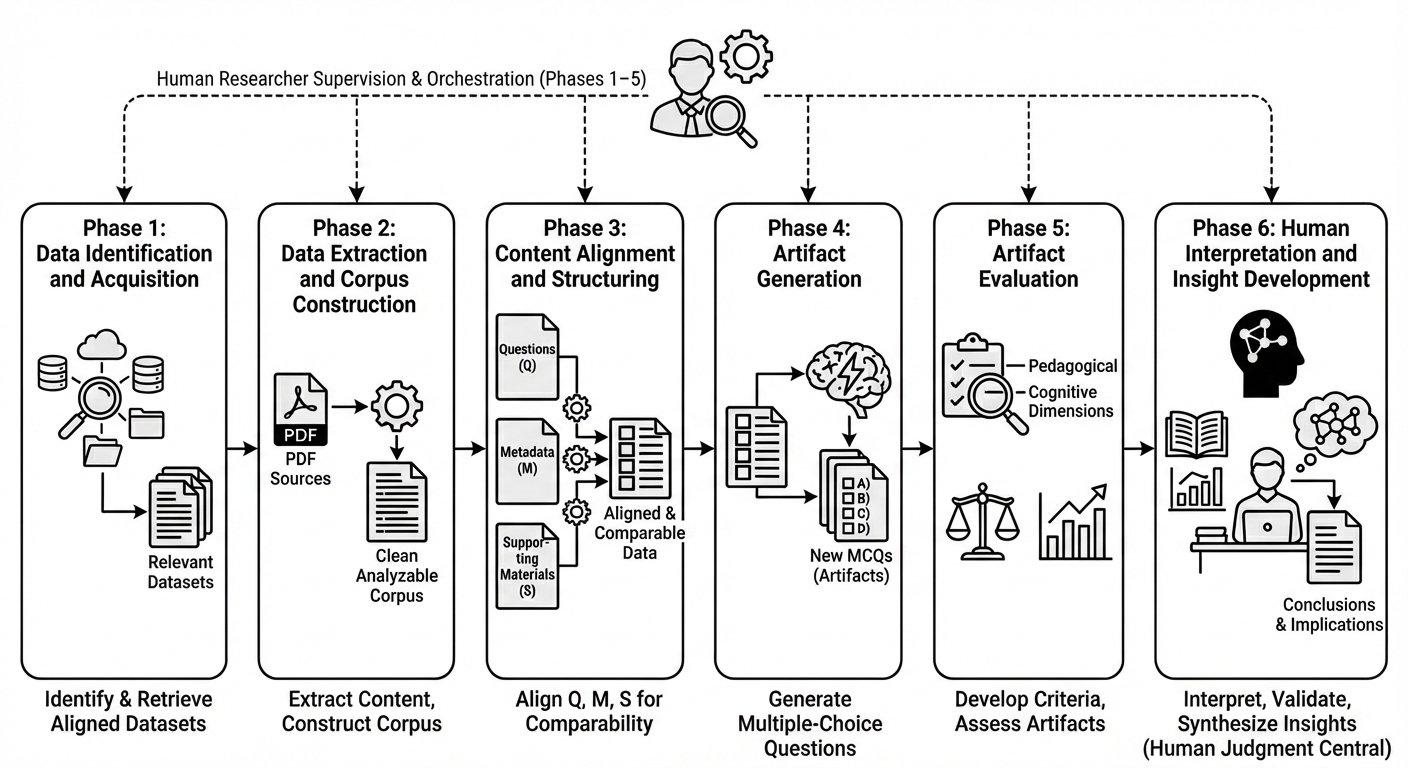}
\caption{Overview of the end-to-end AI-orchestrated MCQ research workflow.}
\label{fig:research-overview}
\end{figure}

Addressing this research question through traditional means would require substantial resources, 
including careful question selection, administration to real students, 
rigorous expert review, and systematic interpretation of results. To reduce cost and improve scalability, 
recent studies have explored the use of LLMs for 
generating educational questions (\cite{biancini2025mcq}), simulating student 
behavior (\cite{martynova-etal-2025-llms,Lu2024GenerativeSU}), modeling student 
responses (\cite{benedetto-etal-2024-using}), and estimating item difficulty (\cite{acquaye2026take}).

Building on this emerging line of work, and given the rapid advancement of AI capabilities, 
we adopt an AI-human collaborative approach. In this framework, a researcher orchestrates a set of
AI tools for data generation, simulation, and preliminary analysis, while the human researcher 
retains responsibility for experimental design, validation, theoretical framing, and 
the interpretation of findings.
Figure \ref{fig:research-overview} details the full architecture of this process.

As the figure illustrates, the study proceeds through the following structured phases:
\begin{itemize}
    \item \textbf{Phase 1: Data Identification and Acquisition:}
Identify and retrieve relevant datasets aligned with the research objectives.

    \item \textbf{Phase 2: Data Extraction and Corpus Construction:}
Extract structured content from PDF sources and construct a clean, analyzable corpus.

    \item \textbf{Phase 3: Content Alignment and Structuring:}
Align questions, metadata, and supporting materials across sources to ensure comparability.

    \item \textbf{Phase 4: Artifact Generation:}
Leverage the aligned corpus to generate new multiple-choice questions (MCQs).

    \item \textbf{Phase 5: Artifact Evaluation:}
Develop evaluation criteria and conduct systematic assessments of generated 
artifacts across pedagogical and cognitive dimensions.

    \item \textbf{Phase 6: Human Interpretation and Insight Development:}
Interpret the data and results, validate findings, and synthesize theoretical and practical insights.
\end{itemize}

Throughout Phases 1-5, the human researcher orchestrates and supervises the AI-driven workflow, 
defining objectives, setting constraints, and monitoring outputs. 
In the final phase, human judgment becomes central, transforming AI-generated 
results into conclusions and implications.

We claim the following contributions in this paper:
\begin{enumerate}
\item We provide an empirical account of an end-to-end 
research workflow executed by a single researcher orchestrating multiple 
LLMs and agentic tools, making  labor shifts from artifact 
production to specification, coordination, and validation.
\item We conduct a large-scale LLM-based evaluation of baseline 
and generated SAT-style MCQs ($\sim$1{,}000 items per set), applying 24 
criteria and two independent LLM judges to examine both average quality 
and criterion-level performance.
\item We demonstrate that despite strong average performance, 
criterion-level equivalence fails on depth and calibration dimensions.
\item We articulate implications for the future of scientific work, 
arguing orchestration, error control, and interpretation as core competencies 
within emerging ``AI research operations.''
\end{enumerate}

The rest of the paper is organized as follows.
Section \ref{sec:related-work} presents the related work.
Section \ref{sec:study_design} describes the study design.
Section \ref{sec:results} reports MCQ quality findings across judges and analysis tracks.
Section \ref{sec:labor-shift} presents work-practice findings.
Section \ref{sec:discussion} discusses the results and their implications.
Section \ref{sec:limitations} provides limitations of the study.
Finally, Section \ref{sec:conclusion} concludes the paper.
The Appendix~\ref{app:ai-tools}-~\ref{app:reproducibility}  contain the tools, 
evaluation criteria, raw data of the study, costs, and the link to the public Github repository.

%%%%%%%%%%%%%%%%%%
%%
%% Related Work
%%
%%%%%%%%%%%%%%%%%%%
\section{Related Work}
\label{sec:related-work}

Our study connects four research areas:
(1) the transformation of scientific work under generative AI;
(2) the structure and automation of scientific workflows; 
(3) the automated generation and evaluation of multiple-choice questions (MCQs); and
(4) the methodological validity and reliability of LLM-as-a-judge 
frameworks for evaluation and assessment.

\subsection{Generative AI and the Future of Knowledge Work}

 \citet{autor2003skill} introduced a taxonomy that
distinguishes routine from non-routine tasks along cognitive and manual dimensions. 
Computerization substitutes for routine cognitive and manual tasks while
complementing non-routine analytic and interactive work. 
\citet{acemoglu2019automation} extend this framework by arguing that 
automation simultaneously displaces labor from
existing tasks and creates entirely new task categories.
Most recently, \citet{dellacqua2023navigating}
provide field-experimental evidence of a ``jagged technological frontier.'' Within
a single occupation, AI dramatically improves performance on some tasks while
degrading it on others, making the task-level boundary between human and machine
competence uneven and hard to predict.

Recent generative-AI work extends these frameworks
to knowledge-intensive tasks by measuring exposure and by running randomized or
quasi-experimental studies. For example, task-exposure analyses suggest broad potential 
impact of LLM-based tooling across occupations, with effects magnified when 
LLMs are embedded in software workflows \citep{eloundou2024gpts}. Controlled studies 
report productivity gains in writing-centric professional tasks when workers 
have access to LLMs \citep{noy2023experimental}, and field evidence from customer support shows 
performance improvements alongside changes in how work is 
executed \citep{brynjolfsson2025generative}.

Most future-of-work evidence to date focuses on relatively bounded tasks.
Scientific work, however, 
is a composite occupation that interleaves information extraction, synthesis, software 
construction, evaluation, and judgment under uncertainty. 
Surveys of ``AI for science'' 
describe an acceleration narrative, from automation toward more autonomous 
agents across the scientific lifecycle, 
but also highlight open problems around verification, bias, and 
governance \citep{zheng2025automation,zhang-etal-2024-comprehensive-survey}. 

\subsection{Scientific Workflows}

Long before LLMs, workflow systems in e-science sought to make computational research 
more reproducible by encoding multi-step pipelines, capturing provenance, and enabling 
re-execution \citep{gil2007examining,deelman2015pegasus}. The literature 
identifies persistent challenges that remain salient in LLM-mediated pipelines.
Managing complex dependencies requires reliability of heterogeneous tools, 
data and metadata quality, provenance capture, 
and the human effort \citep{gil2007examining}.

LLMs alter this landscape by serving as a flexible ``glue'' layer that can generate 
code, call tools, and translate between representations. Work on tool-augmented or 
agentic LLMs shows how models can interleave reasoning with action \citep{yao2022react,schick2023toolformer}. Prompting methods such as 
chain-of-thought can improve reasoning performance but also surface verification 
needs because intermediate rationales may be persuasive yet wrong 
\citep{wei2022chain,kojima2022large}. In software 
engineering, benchmarks such as SWE-bench document both the promise and the limits of 
LLM autonomy in real-world tasks \citep{jimenez223SWE}.

\subsection{Automatic MCQ Generation and Quality}

Automatic question generation (AQG) is a mature research area that predates 
modern LLMs. Surveys of automatic MCQ generation describe modular pipelines 
that identify candidate concepts, construct stems, generate distractors, 
and apply quality filters \citep{ch2018automatic}. 
Parallel work in measurement and classroom assessment provides item-writing 
principles, clarity of the stem, plausibility and homogeneity of distractors, 
avoidance of cueing, and alignment to intended constructs, 
that underpin human-generated test quality \citep{haladyna2002review}.

LLMs change the AQG landscape by making the generation dramatically 
easier, but they also introduce new failure issues such as 
hallucinated facts, inconsistent 
rationales, and superficial distractors.  
Various studies have used MCQ generation as a testbed for evaluating  
LLMs' capabilities on understanding and reasoning
\citep{brown2020language,bommasani2021on,bender2021dangers,achiam2023gpt}. 

Beyond education, AI-driven item generation has been explored in medical
licensing~\citep{kung2023performance} and legal bar
examinations~\citep{katz2024gpt}, where LLMs demonstrate passing-level
performance and, in some cases, generate practice items that experts rate
comparably to human-authored questions. These cross-domain findings
reinforce the generality of the AQG pipeline while underscoring that
domain-specific validity evidence remains essential.

\subsection{LLM-Based Evaluation and ``LLM-as-a-judge''}

Evaluating generative outputs at scale is itself a bottleneck. Recent 
work proposes using strong LLMs as evaluators, reporting substantial 
agreement with human preferences in some settings while also documenting 
systematic biases 
\citep{zheng2023judging,li2025llm,kaiser2025simulating}. 
In natural language generation, frameworks such as G-Eval structure 
LLM judgments via explicit criteria and form-filling, improving alignment 
with human evaluation but still raising concerns about evaluator 
bias toward certain styles or model families \citep{liu-etal-2023-g}.

%%%%%%%%%%%%%%%%%%
%%
%% Context and Design
%%
%%%%%%%%%%%%%%%%%%%
\section{Study Design}
\label{sec:study_design}

This section describes the setting and the end-to-end pipeline.
We document (i) the baseline dataset construction and filtering decisions; (ii) the open-source 
textbook corpus, conversion, and chunking strategy; (iii) the step that links 
each baseline MCQ to a source chunk; (iv) the generation 
of MCQs under explicit metadata constraints; and (v) a LLM-based 
multi-judge evaluation process.

\subsection{Baseline Dataset: SAT Math MCQs}

To establish a reliable baseline for comparison, we prompted an AI system 
to identify authorized SAT question banks. 
It retrieved the official College Board educator question bank~(\cite{collegeboard2026sat}). 
From this source, we focused on SAT Math questions and downloaded the corresponding PDF files, 
which contain 1,682 items across four domains: 
\textit{Algebra}, \textit{Advanced Math}, \textit{Problem Solving and Data Analysis}, 
and \textit{Geometry and Trigonometry}. 

\begin{figure}[htbp]
\centering
\includegraphics[width=\linewidth]{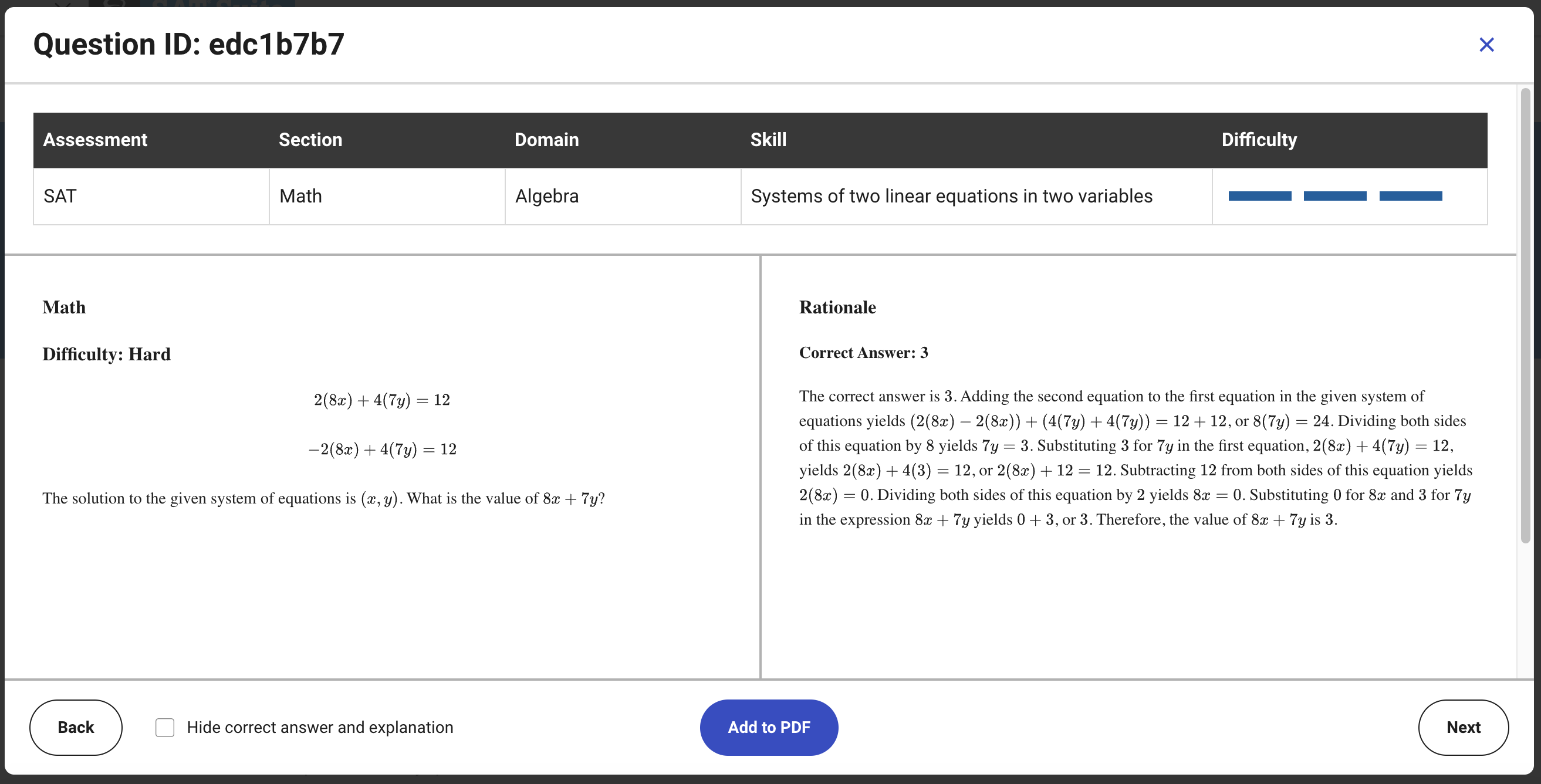}
\caption{Example MCQ entry from the College Board question bank.}
\label{fig:mcq_screenshot}
\end{figure}

We then employed Google AI Studio Gemini 3 Pro Preview  
to convert the PDF content into structured JSON files. 
Our analysis focused exclusively on multiple-choice questions (MCQs). 
Out of the 1,682 SAT Math items, 1,071 were identified as MCQs. 
Each MCQ in the bank has a \texttt{question\_id} and 
associated metadata fields, including 
\texttt{domain}, \texttt{skill}, and \texttt{difficulty}. 
Figure~\ref{fig:mcq_screenshot} shows a screenshot of the details 
of an MCQ in the question bank. During the extraction process, 
we also instructed Gemini 3 Pro to extract a cognitive level, 
\texttt{cognitive\_level\_bloom}, based on Bloom's taxonomy 
\citep{AndersonKrathwohl2001}.

Because some SAT items depend on figures or tables that were not fully captured during extraction, 
we applied an incomplete-item filter to remove partially extracted questions. 
This process resulted in a final corpus of 1,013 SAT Math MCQs filtered by Gemini and
1,027 filtered by GPT for analysis.

\subsection{Open Textbook Corpus and Chunking}
\label{sec:chunking}

The next step is to map each MCQ to source content in textbooks. 
The researcher coordinated AI to search open-source textbooks. It located
the OpenStax (\cite{openstax2026}) which is the world’s largest publisher of open education resources (OER).
Each textbook is downloaded as a PDF file. AI systems were able to convert the PDF file into markdown format, 
and chunk the text into retrievable segments. 
%(Figure~\ref{fig:textbook_chunks}). 
Each chunk is treated as a provenance unit and is referenced by a 
chunk ID during mapping and generation.

%\begin{figure}[htbp]
%centering
%\includegraphics[width=0.5\linewidth]{figures/textbook_chunks.png}
%\caption{Divide textbooks into appropriate chunks for retrieval and provenance. The corpus was 
%split into 2,980 section-based chunks preserving pedagogical coherence.}
%\label{fig:textbook_chunks}
%\end{figure}

We ensure that chunking  preserves pedagogical coherence 
(e.g., keeping a definition with its immediately following example) rather than purely 
token-length-based splitting. We also retained enough surrounding context to 
avoid ``orphan'' equations or references (e.g., ``as shown above'') that would 
otherwise create hallucination pressure during generation.

\subsection{Mapping Baseline MCQs to Chunks}

For each baseline MCQ, we implemented an LLM-assisted retrieval step in
which both MCQs and textbook chunks were embedded into vector representations.
Each MCQ was represented by two queries, a \emph{metadata query} constructed from
the item's domain and skill labels and a \emph{content query} constructed from the question stem and rationale. 
The combined relevance score was computed as:
\[
  \text{score} = 0.3 \times \text{sim}_{\text{metadata}} + 0.7 \times \text{sim}_{\text{content}}
\]
where $\text{sim}$ denotes cosine similarity. 
Using this combined score, the system mapped each MCQ to its most
semantically relevant textbook chunk(s) within the corpus.
These retrieved passages were used as grounding context for
subsequent question generation.

\subsection{Generating MCQ Sets}

To generate new MCQs by LLM from the source material, we first extracted 
the metadata of each baseline MCQ, including its domain, skill, difficulty 
level, and cognitive level. 
We then provided this metadata, together with the retrieved grounding text,
to two representative LLMs, GPT-5-nano (\texttt{gpt-5-nano}, 
accessed February 2026) and Gemini-2.5-Flash (\texttt{gemini-2.5-flash}, accessed 
February 2026),
and prompted them to generate a corresponding MCQ.

\begin{figure}[htbp]
\centering
\includegraphics[width=\linewidth]{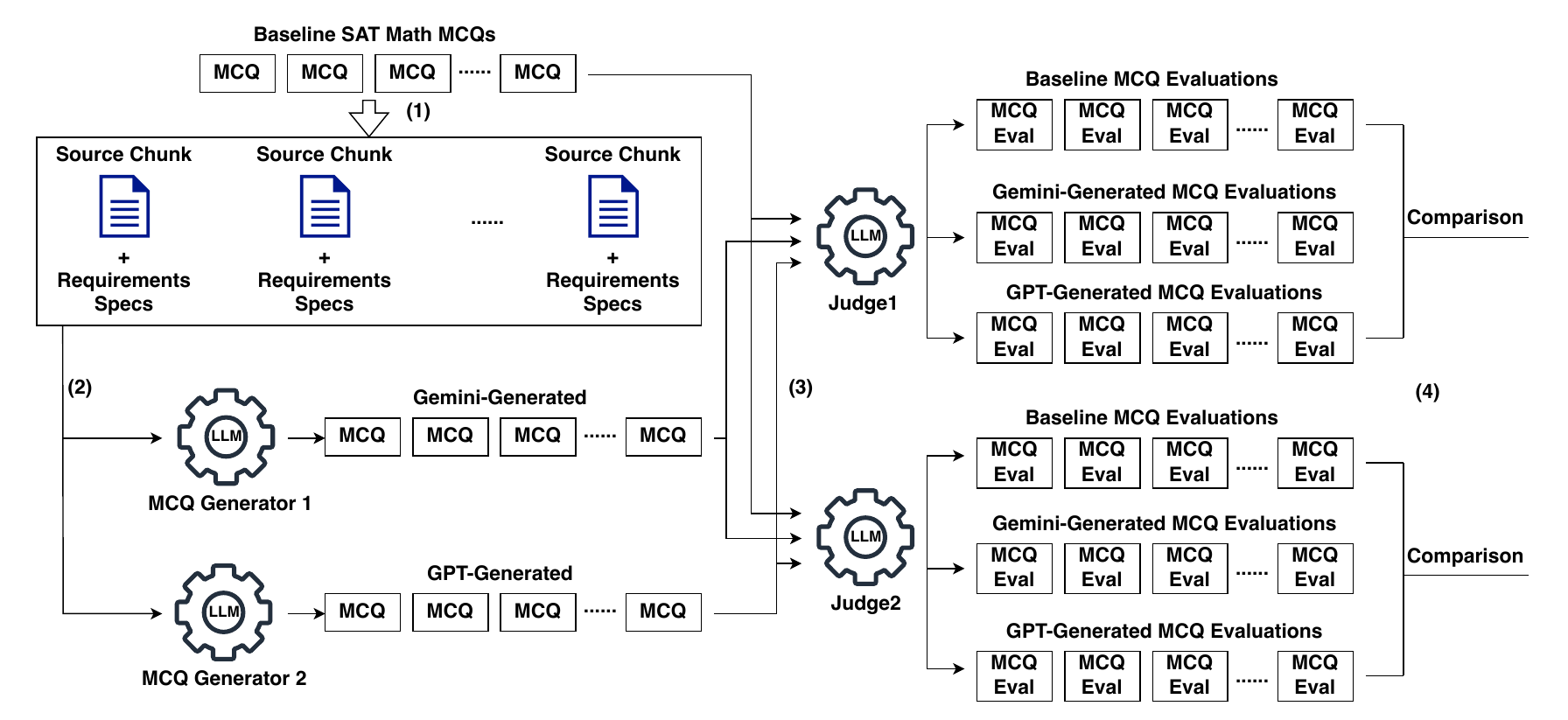}
\caption{LLM-based MCQ generation, evaluation, and comparison}
\label{fig:mcq_generation_evaluation}
\end{figure}

\subsection{Quality Comparison}

\subsubsection{Quality Criteria}
Given the set of metadata along with the mapped source text, an LLM may generate an MCQ that differs 
linguistically from the original MCQ while still satisfying the specified metadata requirements. 
Direct comparison based on surface-level textual similarity is therefore 
neither meaningful nor feasible for assessing alignment with these requirements. 

Our primary interest lies in whether the LLM-generated MCQs match the original 
human-vetted MCQs in overall quality, rather than in word-for-word similarity. 
Accordingly, we prompted an AI system to construct a comprehensive evaluation 
rubric for assessing MCQ quality across multiple dimensions. 
The resulting rubric consists of 24 criteria, each rated on a 
five-point Likert scale with detailed descriptors (see Appendix~\ref{app:rubric}).
The 24 criteria were organized into six thematic  
categories: \emph{Accuracy \& Grounding} (4 criteria), 
\emph{Clarity \& Presentation} (4 criteria), \emph{Distractor \& Answer Quality} (4  
criteria), \emph{Difficulty \& Engagement} (3 criteria), 
\emph{Domain \& Skill Alignment} (5 criteria), and \emph{Difficulty \& Cognitive  
Calibration} (4 criteria).

\subsubsection{Two Independent LLM Judges}
At this stage, we obtained three sets of MCQs: (1) the baseline SAT MCQs, 
(2) MCQs generated by Gemini-2.5-Flash, and 
(3) MCQs generated by GPT-5-nano. 
To evaluate and compare their quality using the established rubric, 
we employed the same two representative LLMs as independent 
evaluators, with each LLM scoring all three sets.
This dual-judge design enables methodological triangulation, allowing us 
to examine potential judge-specific effects and to identify 
evaluation patterns that remain stable across independent model assessors.
Figure \ref{fig:mcq_generation_evaluation} illustrates the MCQ generation and 
judging process.

\subsubsection{Statistical Methods}

Because our central question concerns \emph{comparability}, 
we adopt statistical methods suited for testing practical equivalence. 
Specifically, we employ the two one-sided tests (TOST) procedure, which evaluates 
whether observed differences fall within a pre-specified smallest effect size of interest 
\citep{schuirmann1987comparison}. Equivalence testing has been widely recommended as a 
complement to traditional null-hypothesis significance testing when the goal is to 
determine whether outcomes are ``close enough'' for practical purposes 
\citep{lakens2018equivalence,lakens2017equivalence}. 

We conducted two complementary analysis tracks: 
(i) a \textbf{matched track}, using paired comparisons on overlapping 
\texttt{question\_id}s; and 
(ii) a \textbf{whole-set track}, without matching constraints. 

Equivalence outcomes are summarized as follows:
\begin{itemize}
\item \textbf{Strict similarity:} 24/24 criteria equivalent.
\item \textbf{Practical similarity:} $\geq$19/24 criteria equivalent.
\end{itemize}

%%%%%%%%%%%%%%%%%%
%%
%% Results
%%
%%%%%%%%%%%%%%%%%%%
\section{Results of MCQ Generation and Evaluation}
\label{sec:results}

\noindent This section reports MCQ quality findings at three levels:
(i) aggregated quality;
(ii) criterion-level diagnostics; and
(iii) inferential and equivalence results. 
In the next Section~\ref{sec:labor-shift}, we report work-practice findings.

\subsection{Aggregated Quality}

Mean scores are high in all conditions (Table~\ref{tab:overall-quality}). 
Across judges and MCQ sets, average scores cluster near the top of the 1-5 scale, 
implying that most items are fluent and structurally well-formed. For example:
\begin{itemize}
\item SAT baseline mean: 4.80 (Gemini judge) vs.\ 4.64 (OpenAI GPT judge).
\item Gemini-generated mean: 4.89 (Gemini judge) vs.\ 4.68 (OpenAI GPT judge).
\item GPT-generated mean: 4.82 (Gemini judge) vs.\ 4.66 (OpenAI GPT judge).
\end{itemize}

\begin{table}[h]
\centering
\caption{Overall quality summary (mean over 24 criteria).}
\label{tab:overall-quality}
\small
\begin{tabular}{@{}l|l|l|l|r|r|r@{}}
\toprule
Set & Judge & \# & Mean$\pm$SD & High-quality & Needs review  & Serious  \\
 &  &  &  &  ($=$5) & ($\in$[3, 4]) & ($\in$[1, 2]) \\
\midrule
SAT baseline  & Gemini & 1013 & 4.80 $\pm$ 0.27 & 88\% & 10\% & 2\% \\
Gemini-generated & Gemini & 1065 & 4.89 $\pm$ 0.15 & 93.4\% & 5.2\% & 1.4\% \\
GPT-generated & Gemini & 1071 & 4.82 $\pm$ 0.19 & 90\% & 8\% & 2\% \\
\midrule
SAT baseline & OpenAI & 1016 & 4.64 $\pm$ 0.49 & 74.1\% & 24.5\% & 1.4\% \\
Gemini-generated & OpenAI & 1057 & 4.68 $\pm$ 0.43 & 76\% & 23\% & 1\% \\
GPT-generated & OpenAI & 1058 & 4.66 $\pm$ 0.45 & 75.5\% & 23\% & 1.5\% \\
\bottomrule
\end{tabular}
\end{table}

\paragraph{Interpretation.}
The mean scores of the original SAT MCQs evaluated by both LLM judges are 
lower than the mean scores of the LLM-generated MCQs. 
At face value, these gaps suggest that LLM judges perceive 
LLM-generated MCQs as better calibrated to their internal criteria 
than human-vetted SAT MCQs. Because the origin of an MCQ was not disclosed 
during evaluation, the observed differences likely reflect stylistic 
or structural alignment effects rather than explicit bias based 
on source knowledge. In particular, LLM-generated MCQs may more closely 
match the formatting, phrasing conventions, or implicit preference 
structures embedded in the judging models.

Averages can be misleading. They often mask extreme effects and hide rare but serious failures. Therefore,
we examine criterion-level differences and conduct formal statistical equivalence tests.

\subsection{Criterion-Level Diagnostics}                                          

We conducted evaluation
across all 24 criteria for each of the six judge–generator configurations. 
Table~\ref{tab:criterion-means} reports the mean score per criterion for each configuration.

\begin{table}[htbp]
  \centering
  \caption{Mean evaluation score (1-5) per criterion across six judge-generator configurations. Criteria are grouped by
   thematic category. The three lowest-scoring criteria across all
  configurations are highlighted in \textbf{bold}. G = Gemini judge; O = OpenAI judge; SAT = original SAT items; Gem =
  Gemini-generated; OAI = OpenAI-generated.}
  \label{tab:criterion-means}
  \resizebox{\textwidth}{!}{%
  \begin{tabular}{ll|ccc|ccc}
  \toprule
  \textbf{Category} & \textbf{Criterion} & \textbf{G$\times$SAT} & \textbf{G$\times$Gem} & \textbf{G$\times$OAI} &
  \textbf{O$\times$SAT} & \textbf{O$\times$Gem} & \textbf{O$\times$OAI} \\
   & & $(n{=}1013)$ & $(n{=}1065)$ & $(n{=}1071)$ & $(n{=}1016)$ & $(n{=}1057)$ & $(n{=}1058)$ \\
  \midrule
  \multirow{4}{*}{\rotatebox{90}{\scriptsize Accuracy}}
   & factual\_accuracy         & 4.98 & 4.98 & 4.94 & 4.98 & 4.98 & 4.98 \\
   & no\_hallucinations        & 4.18 & 4.90 & 4.88 & 5.00 & 4.99 & 4.99 \\
   & correct\_key\_answer      & 4.98 & 4.94 & 4.94 & 4.95 & 4.94 & 4.95 \\
   & content\_alignment        & 4.32 & 4.93 & 4.89 & 4.95 & 4.99 & 4.98 \\
  \midrule
  \multirow{4}{*}{\rotatebox{90}{\scriptsize Clarity}}
   & clarity\_question         & 4.99 & 4.98 & 4.97 & 4.87 & 4.94 & 4.95 \\
   & clarity\_options           & 5.00 & 4.98 & 4.98 & 4.96 & 4.94 & 4.95 \\
   & grammar\_fluency          & 4.99 & 5.00 & 5.00 & 4.90 & 4.98 & 4.98 \\
   & no\_duplicates            & 5.00 & 4.98 & 4.98 & 5.00 & 4.98 & 4.98 \\
  \midrule
  \multirow{4}{*}{\rotatebox{90}{\scriptsize Distractor}}
   & no\_answer\_hints         & 5.00 & 5.00 & 4.99 & 4.84 & 4.91 & 4.94 \\
   & \textbf{plausible\_distractors} & \textbf{4.38} & \textbf{4.60} & \textbf{4.49} & \textbf{3.97} & \textbf{4.16} &
  \textbf{4.22} \\
   & unambiguous\_correctness  & 4.97 & 4.93 & 4.93 & 4.96 & 4.92 & 4.95 \\
   & recall\_vs\_guessing      & 4.99 & 4.98 & 4.98 & 4.51 & 4.57 & 4.55 \\
  \midrule
  \multirow{3}{*}{\rotatebox{90}{\scriptsize Diff.\&Eng.}}
   &  {difficulty\_balance} & {4.70} & {4.89} & {4.70} & {3.97} & {4.11} &
  {4.01} \\
   & \textbf{cognitive\_engagement}     & \textbf{4.73} & \textbf{4.88} & \textbf{4.78} & \textbf{3.64} & \textbf{3.78} & \textbf{3.74} \\
   & holistic\_quality         & 4.60 & 4.89 & 4.78 & 4.51 & 4.66 & 4.64 \\
  \midrule
  \multirow{5}{*}{\rotatebox{90}{\scriptsize Skill Align.}}
   & domain\_alignment         & 4.89 & 4.88 & 4.82 & 4.96 & 4.97 & 4.96 \\
   & domain\_appropriateness   & 4.91 & 4.90 & 4.83 & 4.93 & 4.96 & 4.95 \\
   & skill\_alignment          & 4.83 & 4.69 & 4.57 & 4.88 & 4.79 & 4.71 \\
   & \textbf{skill\_depth}     & \textbf{4.18} & \textbf{4.37} & \textbf{4.19} & \textbf{3.22} & \textbf{3.37} &
  \textbf{3.32} \\
   & skill\_isolation          & 4.93 & 4.82 & 4.77 & 4.78 & 4.71 & 4.65 \\
  \midrule
  \multirow{4}{*}{\rotatebox{90}{\scriptsize Calibration}}
   & difficulty\_alignment     & 4.68 & 4.90 & 4.65 & 4.08 & 4.24 & 4.12 \\
   & difficulty\_consistency   & 4.96 & 4.97 & 4.95 & 4.81 & 4.84 & 4.81 \\
   & cog.\_level\_alignment    & 4.96 & 4.98 & 4.86 & 4.83 & 4.85 & 4.74 \\
   & cog.\_level\_approp.      & 4.97 & 4.99 & 4.90 & 4.82 & 4.87 & 4.79 \\
  \midrule
  \multicolumn{2}{l|}{\textbf{Grand mean (24 criteria)}} & \textbf{4.80} & \textbf{4.89} & \textbf{4.82} & \textbf{4.64}
  & \textbf{4.68} & \textbf{4.66} \\
  \bottomrule
  \end{tabular}%
  }
  \end{table}
  
\paragraph{Universally Strong Criteria.}
A cluster of criteria related to surface-level correctness and presentation achieved near-ceiling scores ($>4.9$)
across all six configurations: \texttt{grammar\_fluency} (4.98-5.00), \texttt{no\_duplicates} (4.97-5.00),
\texttt{factual\_accuracy} (4.94-4.98), and \texttt{clarity\_options}
(4.94-4.98). These criteria reflect the ability of both LLM generators (Gemini and GPT) to produce well-formed,
grammatically correct items with non-overlapping answer choices and factually accurate content. The near-perfect
performance on these surface criteria held regardless of judge identity, MCQ source (original SAT vs.\ generated), or
generator model, suggesting that modern LLMs have effectively solved the surface aspects of MCQ authoring.

\paragraph{Baseline Anomaly on Grounding Criteria.}
Two criteria deviate from this pattern in a noteworthy way: \texttt{no\_hallucinations} and \texttt{content\_alignment}
scored substantially lower for the SAT baseline under the Gemini judge (4.18 and 4.32, respectively) than for
either generated set (4.88-4.90 and 4.89-4.93), while the OpenAI judge assigned near-ceiling scores to the
SAT baseline on both criteria (5.00 and 4.95). This asymmetry is best explained as an artifact of the
PDF text extraction process. A subset of SAT items contain references to figures or tables
(e.g., ``according to the graph above'') that were not recoverable from plain-text extraction.
When the Gemini judge evaluates such an item against a textbook chunk that contains no matching
visual, it interprets the unresolved reference as an unsupported claim, triggering a
\texttt{no\_hallucinations} or \texttt{content\_alignment} penalty. LLM-generated items,
by contrast, were written entirely from text-based grounding context and therefore contain no
such unresolvable references. The OpenAI judge, which appears less sensitive to this
pattern, does not penalize the baseline accordingly. This extraction artifact inflates the
apparent quality gap between the SAT baseline and generated sets on these two criteria under the
Gemini judge; both criteria were classified as equivalent under the OpenAI judge
(Table~\ref{tab:criterion-equiv}).

At the category level, \texttt{Accuracy \& Grounding} and \texttt{Clarity \& Presentation} consistently ranked 
as the two strongest
categories across all configurations, with category means of 4.48-4.98 and 4.91-4.99, respectively
(Table~\ref{tab:category-means}). This indicates that the RAG-grounded generation 
pipeline reliably produces MCQs that
are factually anchored to textbook content and clearly articulated.

\begin{table}[h]
  \centering
  \caption{Mean category scores across six judge-generator configurations. Categories are sorted by grand mean across
  all configurations. Scores below 4.5 are underlined.}
  \label{tab:category-means}
  \begin{tabular}{l|ccc|ccc}
  \toprule
  \textbf{Category} & \textbf{G$\times$SAT} & \textbf{G$\times$Gem} & \textbf{G$\times$OAI} & \textbf{O$\times$SAT} &
  \textbf{O$\times$Gem} & \textbf{O$\times$OAI} \\
  \midrule
  Accuracy \& Grounding         & \underline{4.48} & 4.94 & 4.91 & 4.91 & 4.98 & 4.98 \\
  Clarity \& Presentation       & 4.95 & 4.99 & 4.98 & 4.91 & 4.96 & 4.96 \\
  Distractor \& Answer Quality  & 4.73 & 4.88 & 4.85 & 4.54 & 4.64 & 4.66 \\
  Diff.\ \& Cognitive Calibr.   & 4.78 & 4.96 & 4.84 & 4.60 & 4.70 & 4.62 \\
  Domain \& Skill Alignment     & 4.70 & 4.73 & 4.64 & 4.52 & 4.56 & 4.52 \\
  Difficulty \& Engagement      & 4.52 & 4.89 & 4.75 & \underline{4.00} & \underline{4.18} & \underline{4.13} \\
  \bottomrule
  \end{tabular}
\end{table}

\paragraph{Universally Weak Criteria.}
In contrast, a different set of criteria consistently occupied the bottom ranks.
The criterion \texttt{skill\_depth}, which assesses whether a question probes the designated
skill with sufficient rigor rather than superficially invoking it, was the single weakest criterion in all six
configurations. The \texttt{plausible\_distractors} criterion was another 
persistent weakness, with means ranging from 3.97 to 4.60.
\texttt{cognitive\_engagement} (the degree to which an item elicits 
the level of cognitive processing) and \texttt{difficulty\_balance} (appropriate calibration 
of difficulty) also consistently appeared among the bottom five
criteria, with means of 3.64-4.88 and 3.97-4.89, respectively. 

At the category level, \texttt{Domain \& Skill Alignment} (mean: 4.52-4.73) and 
\texttt{Difficulty \& Engagement} (mean: 4.00-4.89) consistently ranked as the two weakest 
categories. The \texttt{Difficulty \& Engagement} category, comprising
\texttt{difficulty\_balance}, \texttt{cognitive\_engagement}, and 
\texttt{holistic\_quality}, showed the largest
spread across configurations, indicating that this dimension 
is both the hardest to get right and the most sensitive to evaluator perspective.

\subsection{Inferential and Equivalence Testing}
\label{sec:equivalence}

For the TOST comparability test,
we set the equivalence bound at $\delta = 0.2\,\textit{SD}$, corresponding to a small
standardized effect \citep{cohen2013statistical}. This threshold was chosen
because differences below 0.2~SD are conventionally regarded as trivially
small. A more stringent bound ($\delta = 0.1$) would
require impractically large samples to detect equivalence, while a more lenient
bound ($\delta = 0.3$) would risk declaring items equivalent. 
A criterion is declared \emph{equivalent}
only when the entire 90\% confidence interval for the mean difference lies within
$[-\delta, +\delta]$.
A criterion was classified as equivalent across item sources when
\emph{all three} pairwise TOST comparisons (Gemini-generated
vs.\ SAT, OpenAI-generated vs.\ SAT, Gemini-generated vs.\
OpenAI-generated) were individually significant at $\alpha=0.05$.

Table~\ref{tab:criterion-equiv} presents the per-criterion
equivalence results.  Eight
criteria were equivalent under every condition (both judges,
both tracks). These consistently equivalent criteria span the core dimensions
of MCQ accuracy, clarity, answer quality, and domain
appropriateness. Conversely, ten criteria were non-equivalent across all
conditions. These criteria predominantly concern calibration against
MCQ metadata targets, including difficulty level, cognitive demand,
and fine-grained skill alignment, as well as distractor
plausibility and overall engagement.
The remaining six criteria showed mixed results that varied
by judge or track, reflecting boundary cases where equivalence
is sensitive to the evaluation perspective.

%% ---------------------------------------------------------------
%% TABLE 1: Criterion-Level Equivalence Verdicts
%% ---------------------------------------------------------------
\begin{table}[!h]
\centering
\caption{Per-criterion equivalence verdicts across judges and
analysis tracks.  A \checkmark\ indicates that all three pairwise
TOST comparisons fell within $\pm 0.2\,\textit{SD}$ at
$\alpha = 0.05$; a \texttimes\ indicates at least one pair was
not equivalent.}
\label{tab:criterion-equiv}
\small
\begin{tabular}{@{}ll cccc@{}}
\toprule
& & \multicolumn{2}{c}{\textbf{Gemini Judge}} &
  \multicolumn{2}{c}{\textbf{OpenAI Judge}} \\
\cmidrule(lr){3-4}\cmidrule(lr){5-6}
\textbf{Category} & \textbf{Criterion} &
  Matched & Whole & Matched & Whole \\
\midrule
\multirow{4}{*}{\makecell[l]{Accuracy \&\\Grounding}}
  & Factual accuracy          & \checkmark & \checkmark & \checkmark & \checkmark \\
  & No hallucinations         & \texttimes & \texttimes & \checkmark & \checkmark \\
  & Correct key answer        & \checkmark & \checkmark & \checkmark & \checkmark \\
  & Content alignment         & \texttimes & \texttimes & \checkmark & \texttimes \\
\midrule
\multirow{4}{*}{\makecell[l]{Clarity \&\\Presentation}}
  & Stem clarity              & \checkmark & \checkmark & \texttimes & \texttimes \\
  & Option clarity            & \checkmark & \checkmark & \checkmark & \checkmark \\
  & Grammar \& fluency        & \checkmark & \checkmark & \texttimes & \texttimes \\
  & No duplicate options      & \checkmark & \checkmark & \checkmark & \checkmark \\
\midrule
\multirow{4}{*}{\makecell[l]{Distractor \&\\Answer Quality}}
  & No answer hints           & \checkmark & \checkmark & \texttimes & \texttimes \\
  & Plausible distractors     & \texttimes & \texttimes & \texttimes & \texttimes \\
  & Unambiguous correctness   & \checkmark & \checkmark & \checkmark & \checkmark \\
  & Recall vs.\ guessing      & \checkmark & \checkmark & \checkmark & \checkmark \\
\midrule
\multirow{3}{*}{\makecell[l]{Difficulty \&\\Engagement}}
  & Difficulty balance        & \texttimes & \texttimes & \texttimes & \texttimes \\
  & Cognitive engagement      & \texttimes & \texttimes & \texttimes & \texttimes \\
  & Holistic quality          & \texttimes & \texttimes & \texttimes & \texttimes \\
\midrule
\multirow{5}{*}{\makecell[l]{Domain \& Skill\\Alignment}}
  & Domain alignment          & \checkmark & \checkmark & \checkmark & \checkmark \\
  & Domain appropriateness    & \checkmark & \texttimes & \checkmark & \texttimes \\
  & Skill alignment           & \texttimes & \texttimes & \texttimes & \texttimes \\
  & Skill depth               & \texttimes & \texttimes & \texttimes & \texttimes \\
  & Skill isolation           & \texttimes & \texttimes & \texttimes & \texttimes \\
\midrule
\multirow{4}{*}{\makecell[l]{Difficulty \&\\Cognitive\\Calibration}}
  & Difficulty alignment      & \texttimes & \texttimes & \texttimes & \texttimes \\
  & Difficulty consistency    & \checkmark & \checkmark & \checkmark & \checkmark \\
  & Cognitive-level alignment & \texttimes & \texttimes & \texttimes & \texttimes \\
  & Cognitive-level approp.   & \texttimes & \texttimes & \texttimes & \texttimes \\
\midrule
\multicolumn{2}{@{}l}{\textbf{Total equivalent (/24)}}
  & \textbf{12} & \textbf{11} & \textbf{11} & \textbf{9} \\
\bottomrule
\end{tabular}
\end{table}

Tables~\ref{tab:category-equiv} and~\ref{tab:category-equiv-whole}
report category-level results.  In the matched track
(Table~\ref{tab:category-equiv}), the Gemini judge found
\texttt{Clarity \& Presentation} and
\texttt{Distractor \& Answer Quality} equivalent (all $|d_z| < 0.08$
for the latter), while the OpenAI judge found
\texttt{Accuracy \& Grounding}, \texttt{Clarity \& Presentation},
and \texttt{Domain \& Skill Alignment} equivalent
(all $|d_z| \leq 0.14$).
Both judges agreed that \texttt{Difficulty \& Engagement}
and \texttt{Difficulty \& Cognitive Calibration}
were not equivalent.
Whole-set results (Table~\ref{tab:category-equiv-whole}) were
largely concordant, with Hedges'~$g$ values corroborating the
matched-track pattern.

%% ---------------------------------------------------------------
%% TABLE 2: Category-Level Equivalence with Effect Sizes
%% ---------------------------------------------------------------
\begin{table}[!h]
\centering
\caption{Category-level equivalence results (matched track).
Each category is the mean of within-criterion $z$-scores for its
constituent criteria.  Cohen's~$d_z$ is reported for each
pairwise comparison; Equiv.\ indicates whether all three pairs
pass TOST at $\delta = 0.2\,\textit{SD}$, $\alpha = 0.05$.
Friedman's $\chi^2$ is the omnibus statistic.}
\label{tab:category-equiv}
\small
\begin{tabular}{@{}l ccc c ccc c@{}}
\toprule
& \multicolumn{4}{c}{\textbf{Gemini Judge} ($n=986$)} &
  \multicolumn{4}{c}{\textbf{OpenAI Judge} ($n=966$)} \\
\cmidrule(lr){2-5}\cmidrule(lr){6-9}
\textbf{Category} &
  \makecell{$d_z$\\G-O} &
  \makecell{$d_z$\\G-S} &
  \makecell{$d_z$\\O-S} &
  Equiv. &
  \makecell{$d_z$\\G-O} &
  \makecell{$d_z$\\G-S} &
  \makecell{$d_z$\\O-S} &
  Equiv. \\
\midrule
Accuracy \& Grounding
  & .057 & .342 & .240 & \texttimes
  & .016 & .044 & .024 & \checkmark \\
Clarity \& Presentation
  & .036 & $-.022$ & $-.055$ & \checkmark
  & $-.019$ & .111 & .136 & \checkmark \\
Distractor \& Answer Qual.
  & .080 & .077 & $-.012$ & \checkmark
  & $-.060$ & .157 & .228 & \texttimes \\
Difficulty \& Engagement
  & .242 & .349 & .097 & \texttimes
  & .090 & .271 & .178 & \texttimes \\
Domain \& Skill Alignment
  & .170 & $-.017$ & $-.141$ & \texttimes
  & .101 & .044 & $-.047$ & \checkmark \\
Difficulty \& Cog.\ Calib.
  & .209 & .134 & $-.106$ & \texttimes
  & .182 & .103 & $-.074$ & \texttimes \\
\midrule
\textbf{Equiv.\ categories}
  & \multicolumn{4}{c}{\textbf{2 / 6}}
  & \multicolumn{4}{c}{\textbf{3 / 6}} \\
\bottomrule
\end{tabular}
\vspace{2pt}

{\footnotesize G = Gemini-generated, O = OpenAI-generated,
S = SAT baseline.}
\end{table}

%% ---------------------------------------------------------------
%% TABLE 3: Whole-Set Track Category-Level Results
%% ---------------------------------------------------------------
\begin{table}[!ht]
\centering
\caption{Category-level equivalence results (whole-set track,
independent samples).  Hedges'~$g$ is reported for each pairwise
comparison.  Welch's ANOVA $F$ provides the omnibus test.}
\label{tab:category-equiv-whole}
\small
\begin{tabular}{@{}l ccc c ccc c@{}}
\toprule
& \multicolumn{4}{c}{\textbf{Gemini Judge}} &
  \multicolumn{4}{c}{\textbf{OpenAI Judge}} \\
\cmidrule(lr){2-5}\cmidrule(lr){6-9}
\textbf{Category} &
  \makecell{$g$\\G-O} &
  \makecell{$g$\\G-S} &
  \makecell{$g$\\O-S} &
  Equiv. &
  \makecell{$g$\\G-O} &
  \makecell{$g$\\G-S} &
  \makecell{$g$\\O-S} &
  Equiv. \\
\midrule
Accuracy \& Grounding
  & .058 & .472 & .369 & \texttimes
  & .012 & .052 & .036 & \checkmark \\
Clarity \& Presentation
  & .036 & $-.031$ & $-.071$ & \checkmark
  & $-.036$ & .137 & .182 & \texttimes \\
Distractor \& Answer Qual.
  & .084 & .087 & $-.006$ & \checkmark
  & $-.080$ & .209 & .306 & \texttimes \\
Difficulty \& Engagement
  & .329 & .497 & .142 & \texttimes
  & .118 & .349 & .232 & \texttimes \\
Domain \& Skill Alignment
  & .131 & $-.040$ & $-.179$ & \texttimes
  & .117 & .055 & $-.069$ & \checkmark \\
Difficulty \& Cog.\ Calib.
  & .296 & .195 & $-.144$ & \texttimes
  & .217 & .142 & $-.072$ & \texttimes \\
\midrule
\textbf{Equiv.\ categories}
  & \multicolumn{4}{c}{\textbf{2 / 6}}
  & \multicolumn{4}{c}{\textbf{2 / 6}} \\
\bottomrule
\end{tabular}
\vspace{2pt}

{\footnotesize G = Gemini-generated
($n_{\mathrm{Gem}} \approx 1{,}060$),
O = OpenAI-generated ($n_{\mathrm{OAI}} \approx 1{,}070$),
S = SAT baseline ($n_{\mathrm{SAT}} \approx 1{,}015$).}
\end{table}

Table~\ref{tab:equiv-summary} summarises the overall outcomes.
Track agreement was high (95.8\% for the Gemini judge, 91.7\%
for the OpenAI judge).

%% ---------------------------------------------------------------
%% TABLE 4: Cross-Condition Agreement Summary
%% ---------------------------------------------------------------
\begin{table}[!ht]
\centering
\caption{Summary of equivalence testing outcomes across all
conditions.  Track agreement indicates the proportion of criteria
whose equivalence verdict is the same under both analysis tracks
for a given judge.}
\label{tab:equiv-summary}
\begin{tabular}{@{}lcccc@{}}
\toprule
& \multicolumn{2}{c}{\textbf{Gemini Judge}} &
  \multicolumn{2}{c}{\textbf{OpenAI Judge}} \\
\cmidrule(lr){2-3}\cmidrule(lr){4-5}
\textbf{Metric} & Matched & Whole & Matched & Whole \\
\midrule
Equivalent criteria (/24)
  & 12 & 11 & 11 & 9 \\
Equivalent categories (/6)
  & 2 & 2 & 3 & 2 \\
Track agreement (criteria)
  & \multicolumn{2}{c}{23/24 (95.8\%)}
  & \multicolumn{2}{c}{22/24 (91.7\%)} \\
\bottomrule
\end{tabular}
\end{table}

\paragraph{Interpretation.}
Overall, the generated sets show \emph{partial} comparability to baseline: a substantial subset
of criteria meets equivalence, but strict all-criteria comparability is not supported in either
evaluator or analysis track.

\subsection{Judge Effects}

Gemini-judged scores are consistently higher than OpenAI-judged scores by roughly 0.16--0.21 points 
in mean (Table~\ref{tab:overall-quality}), indicating a rater effect that can be comparable to condition effects. 
In measurement terms, the choice of evaluator changes the observed scale location and may alter which criteria appear problematic.

This finding supports two operational recommendations. First, evaluations should be treated as a 
governed measurement process (with documentation, calibration checks, and sensitivity analyses), 
not as an invisible backend step. Second, conclusions about ``quality'' should be reported with evaluator 
provenance (which judge model, prompt, and scoring procedure), because these choices are part of the research instrument.

%%%%%%%%%%%%%%%%%%
%%
%% Work-Practice Findings
%%
%%%%%%%%%%%%%%%%%%%
\section{Work-Practice Findings}
\label{sec:labor-shift}

The project was completed over 10 days, with approximately 80 hours and LLM costs  
\$62. Appendix~\ref{app:cost} lists the cost breakdown.

\subsection{Task Classification}

To characterize the labor shift concretely, we classified all research activities
(Appendix~\ref{app:activity_logs}) during the 10-day project into five task categories, 
drawing on the
routine/non-routine distinction of \citet{autor2003skill}.
Table~\ref{tab:task-classification} maps each category to concrete task examples
from the project and indicates whether the task is routine-cognitive (automatable
by LLMs) or non-routine analytic (requiring human judgment).

\begin{table}[htbp]
\centering
\caption{Classification of research tasks in the AI-orchestrated workflow,
mapped to the task framework of \citet{autor2003skill}.}
\label{tab:task-classification}
\small
\begin{tabular}{@{}p{3.2cm}p{5.5cm}p{3.8cm}@{}}
\toprule
\textbf{Task Category} & \textbf{Concrete Examples} & \textbf{Autor et al.\ Type} \\
\midrule
Research Framing and Problem Definition
  & Day 1: defining the core hypothesis and comparison design; deciding 
  baseline and success condition
  & Non-routine analytic \\
\addlinespace
Data Acquisition and Formatting
  & Day 1-2: downloading SAT PDFs; extracting PDF content to JSON; 
  extracting MCQs; manual source collection from OpenStax
  & Mixed: non-routine: prompting AIs, manually download PDFs; 
  routine: generating code for file conversion \\
\addlinespace
Pipeline Engineering and Operations
  & Day 3-8: building/testing scripts for OCR batching, JSON merging, 
  chunking, embedding, retrieval mapping, segmented generation runs
  & Mixed: routine-heavy with non-routine troubleshooting \\
\addlinespace
Evaluation Design and Statistical Analysis
  & Day 9-10: expanding criteria 15 $\rightarrow$ 24; implementing multi-set evaluation 
  pipeline; aggregation and formal comparability analysis
  & Mixed: non-routine: analytics; routine: generate the analytic programs \\
\addlinespace
Human Quality Control and Interpretation
  & Day 4-10: validating scripts/results, spot-checking outputs, 
  agreeing/disagreeing with model ratings, interpreting final findings
  & Non-routine: interaction/judgment \\
\bottomrule
\end{tabular}
\end{table}

The routine cognitive tasks, data
extraction, item formatting, rubric-based scoring, and code generation, were
delegated to LLM agents. The researcher's residual work was
entirely composed of tasks requiring judgment, design, and system-level reasoning.
This pattern is consistent with \citet{acemoglu2019automation}'s observation
that automation does not merely eliminate tasks but creates new ones:
orchestration, verification, and governance are emergent task categories
that did not exist in the traditional item-writing workflow.

\subsection{Time Allocation}

Table~\ref{tab:time-allocation} reports the approximate distribution of
researcher time across the five task categories during the 10-day project. 

\begin{table}[htbp]
\centering
\caption{Approximate researcher time allocation across task categories
(total $\approx$80 hours over 10 days).}
\label{tab:time-allocation}
\begin{tabular}{@{}lrr@{}}
\toprule
\textbf{Task Category} & \textbf{Hours} & \textbf{\%} \\
\midrule
Research Framing and Problem Definition         & 8  & 10\% \\
Data Acquisition and Formatting    & 10  & 12.5\% \\
Pipeline Engineering and Operations              & 42  & 52.5\% \\
Evaluation Design and Statistical Analysis  &  12  & 15\% \\
Human Quality Control and Interpretation     &  8  & 10\% \\
\midrule
\textbf{Total}                  & \textbf{80} & \textbf{100\%} \\
\bottomrule
\end{tabular}
\end{table}

The largest share of researcher time (52.5\%) went to 
Pipeline Engineering and Operations, 
rather than to content creation,
which was fully delegated to LLM agents. 
Research framing, evaluation design, and human interpretation
together accounted for another 35\%, reinforcing the finding that
the researcher's role shifted from \emph{producing} artifacts to
\emph{defining and controlling} the production process.

\subsection{Efficiency Comparison}

For a researcher who already possesses the required domain and tooling expertise,
the AI-orchestrated workflow produced a striking compression of elapsed time.
A project of this scope, encompassing data ingestion, source mapping,
question synthesis, and evaluation framework design, would typically require
approximately six months of full-time doctoral student effort.
Table~\ref{tab:labor-comparison} presents this as an illustrative contrast.

\begin{table}[htbp]
\centering
\caption{\textbf{Illustrative contrast of research resource requirements.}}
\label{tab:labor-comparison}
\begin{tabular}{@{}lll@{}}
\toprule
\textbf{Metric} & \textbf{Traditional PhD Timeline} & \textbf{AI-Orchestrated Workflow} \\ \midrule
Estimated Duration & $\sim$6 Months & $\sim$10 Days \\
Labor Intensity & Full-time & $\sim$8 Hours/Day \\
Variable API Cost & --- & $\sim$\$62 (Mistral + OpenAI + Gemini) \\
\bottomrule
\end{tabular}
\end{table}

The six-month estimate reflects the typical pace of a doctoral researcher
performing the constituent tasks manually with writing program code. 
The timeline is a retrospective estimate; it is not based on empirical measurement.
The tasks and time are roughly estimated as: literature search and dataset
identification ($\sim$2 weeks), PDF extraction and data cleaning ($\sim$4 weeks),
textbook corpus construction and alignment ($\sim$4 weeks),  
MCQ generation and review ($\sim$8 weeks), evaluation framework design and scoring ($\sim$4 weeks),
and statistical analysis and write-up ($\sim$4 weeks). These estimates are
conservative and exclude the learning curve for domain expertise.

In Table~\ref{tab:labor-comparison}, 
the two conditions differ not only in duration but also in labor quality, 
in output quality assurance, and in the cost structure.
Both conditions exclude researcher salary. The AI-Orchestrated
column also excludes subscription services (\$40/month for Claude Code and
Codex) as fixed costs. The direct cost figure reflects variable API charges
only and should not be interpreted as the total cost of the research.

The appropriate interpretation is not that AI reduces research cost, 
but that it dramatically compresses elapsed \emph{time} for a skilled
practitioner who can translate research goals into reliable pipeline specifications.
This time compression is itself consequential. It changes what is feasible within
a grant cycle, a semester, or a conference deadline.

\subsection{Deskilling or Reskilling}

A natural question is whether the shift from item authoring to orchestration
represents a \emph{deskilling} or \emph{reskilling} of the researcher role.
Our experience suggests the latter. Orchestration work requires a distinctive
skill set that combines domain expertise,
software engineering judgment, and evaluation
literacy. These competencies
are not typically part of doctoral training, suggesting that ``AI research operations'' may
emerge as a recognized professional capability.

This observation connects to the reskilling theme: as LLMs
absorb content-generation tasks, researchers need training not in prompt
engineering alone, but in \emph{specification design}, \emph{failure-mode
analysis}, and \emph{evaluation governance}, skills that are closer to
systems engineering than to traditional item writing. This pattern is
not without precedent. The emergence of bioinformatics in the 1990s
created a hybrid role that combined biological domain knowledge with
computational pipeline expertise, a skill set that did not exist as a
recognized specialty before sequencing data made it necessary
\citep{gauthier2019brief}. Similarly, ``data scientist'' emerged as
a recognized professional role as statistical analysis was embedded in
scalable software pipelines, requiring researchers to master both domain
interpretation and engineering infrastructure. ``AI research operations''
may follow the same trajectory.

%%%%%%%%%%%%%%%%%%
%%
%% Discussion
%%
%%%%%%%%%%%%%%%%%%%
\section{Discussion}
\label{sec:discussion}

This study suggests that the real gains from LLMs are not coming 
from ``hands-free'' automation, but from a reshaping of the work itself. 
In our workflow, the models took on much of the discovering resources, 
generating code, along with initial evaluation. But that did not eliminate 
human effort. We had to translate broad goals into precise specifications, 
define constraints, build checks for failure cases, and maintain clear 
boundaries. As generating content became easier, 
the harder problem became controlling the workflow. The work did 
not vanish; it shifted toward design, coordination, and quality control.

We also found that different LLM-based judges produced systematically
different results. That means evaluation choices can shape conclusions in meaningful ways.
Responsible adoption, therefore, requires explicit governance. It consists of 
documenting how items are evaluated, testing sensitivity to different judges, defining
failure thresholds, and logging model and prompt
versions. Using AI in research is less about replacing people
and more about building systems that can reliably specify,
monitor, and audit high-volume machine-generated work.

%%%%%%%%%%%%%%%%%%
%%
%% Limitations
%%
%%%%%%%%%%%%%%%%%%%
\section{Limitations }
\label{sec:limitations}

This study has several limitations that should be considered when
interpreting the results.

\paragraph{Single researcher in a single domain.}
The workflow was executed by a single researcher in a single domain
(SAT Math). The observed labor shift and efficiency gains may not
generalize to team-based research settings, where coordination
overhead, knowledge distribution, and division of labor introduce
additional complexity. Similarly, the MCQ generation and evaluation
results are specific to mathematics; domains requiring visual reasoning,
laboratory procedures, or open-ended responses may present different
challenges for LLM-based pipelines.

\paragraph{LLM-as-judge.}
A key concern is that LLMs are used to both generate and evaluate 
artifacts based on a set of criteria. 
Although our dual-judge design provides 
partial mitigation, both judges are LLMs and may share systematic blind spots. 
Stronger validation would include human expert raters and 
inter-rater reliability analysis. In the future, we will
conduct human verification with the LLM-evaluation results and
evaluate whether stronger LLM models produce more reliable justifications.

Another possible extension would be to use an LLM to simulate a 
classroom of students answering MCQs. 
Such a simulation could generate empirical difficulty 
estimates (e.g., p-values and Rasch parameters) based on 
item response theory (IRT) modeling~\citep{hambleton_item_2013}. 
We reserve this direction for follow-up work.

\paragraph{AI-generated rubric provenance.}
The 24-criterion evaluation rubric was itself developed with AI
assistance, raising a provenance concern. The evaluation instrument
may implicitly favor the kinds of quality dimensions that LLMs are
good at detecting while underweighting dimensions that require human
pedagogical expertise. 
Future iterations should involve domain experts in rubric design and validation.

\paragraph{Generalizability.}
The efficiency comparison (Table~\ref{tab:labor-comparison}) is based
on a single project and should not be extrapolated to other research
contexts without careful consideration of domain complexity, researcher
expertise, and tooling maturity. The comparison also excludes the
researcher's own time cost, which would reduce the apparent
efficiency advantage.

%%%%%%%%%%%%%%%%%%
%%
%% Conclusion
%%
%%%%%%%%%%%%%%%%%%%
\section{Conclusion}
\label{sec:conclusion}

This pilot provides empirical evidence that 
AI-orchestrated research workflows can scale end-to-end research 
operations while shifting human labor toward orchestration and verification. 
Using SAT Math MCQ generation and evaluation as a testbed, we 
showed that a single researcher coordinating multiple LLMs and agentic 
tools can produce and evaluate large volumes of artifacts under explicit 
constraints and with traceable provenance.

At the workflow level, the study documents a qualitative labor shift 
consistent with future-of-work frameworks. As execution tasks become 
cheaper to automate, the scarce and consequential work moves to system-level 
control, including specifying constraints, orchestrating heterogeneous tools, verifying 
correctness and alignment, and 
maintaining auditable records. In this sense, the primary value of LLMs in
scientific work is not simply faster writing or coding, but the
enabling of new forms of high-throughput
research.

As a pilot study, this work establishes feasibility and identifies key
patterns, but a full validation would require several additional steps:
(i)~independent human expert raters scoring a representative subset of
items to anchor LLM judge outputs;
(ii)~administration of generated items to real students, with
item-response-theory (IRT) analysis to estimate psychometric properties;
(iii)~multi-researcher replication to assess whether the observed labor
shift and efficiency gains hold across different skill levels and
working styles; and
(iv)~extension to other domains and item types beyond SAT Math MCQs.
These directions would move the evidence base from proof-of-concept
toward operational readiness.

%\newpage
%\bibliography{references/references,references/references_IUSE2026}

\begin{thebibliography}{43}
\providecommand{\natexlab}[1]{#1}
\providecommand{\url}[1]{\texttt{#1}}
\expandafter\ifx\csname urlstyle\endcsname\relax
  \providecommand{\doi}[1]{doi: #1}\else
  \providecommand{\doi}{doi: \begingroup \urlstyle{rm}\Url}\fi

\bibitem[Acemoglu and Restrepo(2019)]{acemoglu2019automation}
Daron Acemoglu and Pascual Restrepo.
\newblock Automation and new tasks: How technology displaces and reinstates
  labor.
\newblock \emph{Journal of economic perspectives}, 33\penalty0 (2):\penalty0
  3--30, 2019.

\bibitem[Achiam et~al.(2023)Achiam, Adler, Agarwal, Ahmad, Akkaya, Aleman,
  Almeida, Altenschmidt, Altman, Anadkat, et~al.]{achiam2023gpt}
Josh Achiam, Steven Adler, Sandhini Agarwal, Lama Ahmad, Ilge Akkaya,
  Florencia~Leoni Aleman, Diogo Almeida, Janko Altenschmidt, Sam Altman,
  Shyamal Anadkat, et~al.
\newblock Gpt-4 technical report.
\newblock \emph{arXiv preprint arXiv:2303.08774}, 2023.

\bibitem[{Acquaye} et~al.(2026){Acquaye}, {Huang}, {Carpuat}, and
  {Rudinger}]{acquaye2026take}
Christabel {Acquaye}, Yi~Ting {Huang}, Marine {Carpuat}, and Rachel {Rudinger}.
\newblock {Take Out Your Calculators: Estimating the Real Difficulty of
  Question Items with LLM Student Simulations}.
\newblock \emph{arXiv e-prints}, art. arXiv:2601.09953, January 2026.
\newblock \doi{10.48550/arXiv.2601.09953}.

\bibitem[Anderson and Krathwohl(2001)]{AndersonKrathwohl2001}
Lorin~W. Anderson and David~R. Krathwohl, editors.
\newblock \emph{A Taxonomy for Learning, Teaching, and Assessing: A Revision of
  Bloom's Taxonomy of Educational Objectives}.
\newblock Longman, New York, 2001.
\newblock ISBN 978-0-8013-1903-7.

\bibitem[Autor et~al.(2003)Autor, Levy, and Murnane]{autor2003skill}
David~H. Autor, Frank Levy, and Richard~J. Murnane.
\newblock The skill content of recent technological change: An empirical
  exploration.
\newblock \emph{The Quarterly Journal of Economics}, 118\penalty0 (4):\penalty0
  1279--1333, 2003.

\bibitem[Bender et~al.(2021)Bender, Gebru, McMillan-Major, and
  Shmitchell]{bender2021dangers}
Emily~M Bender, Timnit Gebru, Angelina McMillan-Major, and Shmargaret
  Shmitchell.
\newblock On the dangers of stochastic parrots: Can language models be too big?
\newblock In \emph{Proceedings of the 2021 ACM conference on fairness,
  accountability, and transparency}, pages 610--623, 2021.

\bibitem[Benedetto et~al.(2024)Benedetto, Aradelli, Donvito, Lucchetti,
  Cappelli, and Buttery]{benedetto-etal-2024-using}
Luca Benedetto, Giovanni Aradelli, Antonia Donvito, Alberto Lucchetti, Andrea
  Cappelli, and Paula Buttery.
\newblock Using {LLM}s to simulate students' responses to exam questions.
\newblock In Yaser Al-Onaizan, Mohit Bansal, and Yun-Nung Chen, editors,
  \emph{Findings of the Association for Computational Linguistics: EMNLP 2024},
  pages 11351--11368, Miami, Florida, USA, November 2024. Association for
  Computational Linguistics.
\newblock \doi{10.18653/v1/2024.findings-emnlp.663}.
\newblock URL \url{https://aclanthology.org/2024.findings-emnlp.663/}.

\bibitem[{Biancini} et~al.(2025){Biancini}, {Ferrato}, and
  {Limongelli}]{biancini2025mcq}
Giorgio {Biancini}, Alessio {Ferrato}, and Carla {Limongelli}.
\newblock {Multiple-Choice Question Generation Using Large Language Models:
  Methodology and Educator Insights}.
\newblock \emph{arXiv e-prints}, art. arXiv:2506.04851, June 2025.
\newblock \doi{10.48550/arXiv.2506.04851}.

\bibitem[{Bommasani} et~al.(2021){Bommasani}, {Hudson},
  et~al.]{bommasani2021on}
Rishi {Bommasani}, Drew~A. {Hudson}, et~al.
\newblock {On the Opportunities and Risks of Foundation Models}.
\newblock \emph{arXiv e-prints}, art. arXiv:2108.07258, August 2021.
\newblock \doi{10.48550/arXiv.2108.07258}.

\bibitem[Brown et~al.(2020)Brown, Mann, Ryder, Subbiah, Kaplan, Dhariwal,
  Neelakantan, Shyam, Sastry, Askell, et~al.]{brown2020language}
Tom Brown, Benjamin Mann, Nick Ryder, Melanie Subbiah, Jared~D Kaplan, Prafulla
  Dhariwal, Arvind Neelakantan, Pranav Shyam, Girish Sastry, Amanda Askell,
  et~al.
\newblock Language models are few-shot learners.
\newblock \emph{Advances in neural information processing systems},
  33:\penalty0 1877--1901, 2020.

\bibitem[Brynjolfsson et~al.(2025)Brynjolfsson, Li, and
  Raymond]{brynjolfsson2025generative}
Erik Brynjolfsson, Danielle Li, and Lindsey Raymond.
\newblock Generative ai at work.
\newblock \emph{The Quarterly Journal of Economics}, 140\penalty0 (2):\penalty0
  889--942, 2025.

\bibitem[CH and Saha(2020)]{ch2018automatic}
Dhawaleswar~Rao CH and Sujan~Kumar Saha.
\newblock Automatic multiple choice question generation from text: A survey.
\newblock \emph{IEEE Transactions on Learning Technologies}, 13\penalty0
  (1):\penalty0 14--25, 2020.
\newblock \doi{10.1109/TLT.2018.2889100}.

\bibitem[Cohen(2013)]{cohen2013statistical}
Jacob Cohen.
\newblock \emph{Statistical power analysis for the behavioral sciences}.
\newblock routledge, 2013.

\bibitem[{College Board}(2026)]{collegeboard2026sat}
{College Board}.
\newblock Sat suite educator question bank, 2026.
\newblock URL \url{https://satsuiteeducatorquestionbank.collegeboard.org/}.
\newblock Accessed: 2026-02-13.

\bibitem[Deelman et~al.(2015)Deelman, Vahi, Juve, Rynge, Callaghan, Maechling,
  Mayani, Chen, Da~Silva, Livny, et~al.]{deelman2015pegasus}
Ewa Deelman, Karan Vahi, Gideon Juve, Mats Rynge, Scott Callaghan, Philip~J
  Maechling, Rajiv Mayani, Weiwei Chen, Rafael~Ferreira Da~Silva, Miron Livny,
  et~al.
\newblock Pegasus, a workflow management system for science automation.
\newblock \emph{Future Generation Computer Systems}, 46:\penalty0 17--35, 2015.

\bibitem[Dell'Acqua et~al.(2023)Dell'Acqua, McFowland~III, Mollick,
  Lifshitz-Assaf, Kellogg, Rajendran, Krayer, Candelon, and
  Lakhani]{dellacqua2023navigating}
Fabrizio Dell'Acqua, Edward McFowland~III, Ethan~R. Mollick, Hila
  Lifshitz-Assaf, Katherine Kellogg, Saran Rajendran, Lisa Krayer,
  Fran{\c{c}}ois Candelon, and Karim~R. Lakhani.
\newblock Navigating the jagged technological frontier: Field experimental
  evidence of the effects of {AI} on knowledge worker productivity and quality.
\newblock \emph{Harvard Business School Technology \& Operations Mgt.\ Unit
  Working Paper}, \penalty0 (24-013), 2023.

\bibitem[Eloundou et~al.(2024)Eloundou, Manning, Mishkin, and
  Rock]{eloundou2024gpts}
Tyna Eloundou, Sam Manning, Pamela Mishkin, and Daniel Rock.
\newblock Gpts are gpts: Labor market impact potential of llms.
\newblock \emph{Science}, 384\penalty0 (6702):\penalty0 1306--1308, 2024.

\bibitem[Gauthier et~al.(2019)Gauthier, Vincent, Charette, and
  Derome]{gauthier2019brief}
Jeff Gauthier, Antony~T Vincent, Steve~J Charette, and Nicolas Derome.
\newblock A brief history of bioinformatics.
\newblock \emph{Briefings in bioinformatics}, 20\penalty0 (6):\penalty0
  1981--1996, 2019.

\bibitem[Gil et~al.(2007)Gil, Deelman, Ellisman, Fahringer, Fox, Gannon, Goble,
  Livny, Moreau, and Myers]{gil2007examining}
Yolanda Gil, Ewa Deelman, Mark Ellisman, Thomas Fahringer, Geoffrey Fox, Dennis
  Gannon, Carole Goble, Miron Livny, Luc Moreau, and Jim Myers.
\newblock Examining the challenges of scientific workflows.
\newblock \emph{Computer}, 40\penalty0 (12):\penalty0 24--32, 2007.

\bibitem[Haladyna et~al.(2002)Haladyna, Downing, and
  Rodriguez]{haladyna2002review}
Thomas~M Haladyna, Steven~M Downing, and Michael~C Rodriguez.
\newblock A review of multiple-choice item-writing guidelines for classroom
  assessment.
\newblock \emph{Applied measurement in education}, 15\penalty0 (3):\penalty0
  309--333, 2002.

\bibitem[Hambleton and Swaminathan(2013)]{hambleton_item_2013}
Ronald~K. Hambleton and Hariharan Swaminathan.
\newblock \emph{Item response theory: {Principles} and applications}.
\newblock Springer Science \& Business Media, 2013.
\newblock ISBN 94-017-1988-8.

\bibitem[{Jimenez} et~al.(2023){Jimenez}, {Yang}, {Wettig}, {Yao}, {Pei},
  {Press}, and {Narasimhan}]{jimenez223SWE}
Carlos~E. {Jimenez}, John {Yang}, Alexander {Wettig}, Shunyu {Yao}, Kexin
  {Pei}, Ofir {Press}, and Karthik {Narasimhan}.
\newblock {SWE-bench: Can Language Models Resolve Real-World GitHub Issues?}
\newblock \emph{arXiv e-prints}, art. arXiv:2310.06770, October 2023.
\newblock \doi{10.48550/arXiv.2310.06770}.

\bibitem[Katz et~al.(2024)Katz, Bommarito, Gao, and Arredondo]{katz2024gpt}
Daniel~Martin Katz, Michael~James Bommarito, Shang Gao, and Pablo Arredondo.
\newblock Gpt-4 passes the bar exam.
\newblock \emph{Philosophical Transactions of the Royal Society A:
  Mathematical, Physical and Engineering Sciences}, 382\penalty0 (2270), 2024.

\bibitem[Kojima et~al.(2022)Kojima, Gu, Reid, Matsuo, and
  Iwasawa]{kojima2022large}
Takeshi Kojima, Shixiang~Shane Gu, Machel Reid, Yutaka Matsuo, and Yusuke
  Iwasawa.
\newblock Large language models are zero-shot reasoners.
\newblock \emph{Advances in neural information processing systems},
  35:\penalty0 22199--22213, 2022.

\bibitem[Kung et~al.(2023)Kung, Cheatham, Medenilla, Sillos, De~Leon,
  Elepa{\~n}o, Madriaga, Aggabao, Diaz-Candido, Maningo,
  et~al.]{kung2023performance}
Tiffany~H Kung, Morgan Cheatham, Arielle Medenilla, Czarina Sillos, Lorie
  De~Leon, Camille Elepa{\~n}o, Maria Madriaga, Rimel Aggabao, Giezel
  Diaz-Candido, James Maningo, et~al.
\newblock Performance of chatgpt on usmle: potential for ai-assisted medical
  education using large language models.
\newblock \emph{PLoS digital health}, 2\penalty0 (2):\penalty0 e0000198, 2023.

\bibitem[Lakens(2017)]{lakens2017equivalence}
Dani{\"e}l Lakens.
\newblock Equivalence tests: A practical primer for t tests, correlations, and
  meta-analyses.
\newblock \emph{Social psychological and personality science}, 8\penalty0
  (4):\penalty0 355--362, 2017.

\bibitem[Lakens et~al.(2018)Lakens, Scheel, and Isager]{lakens2018equivalence}
Dani{\"e}l Lakens, Anne~M Scheel, and Peder~M Isager.
\newblock Equivalence testing for psychological research: A tutorial.
\newblock \emph{Advances in methods and practices in psychological science},
  1\penalty0 (2):\penalty0 259--269, 2018.

\bibitem[Li et~al.(2025)Li, Chen, Namkoong, and Peng]{li2025llm}
Ang Li, Haozhe Chen, Hongseok Namkoong, and Tianyi Peng.
\newblock Llm generated persona is a promise with a catch.
\newblock \emph{arXiv preprint arXiv:2503.16527}, 2025.

\bibitem[Liu et~al.(2023)Liu, Iter, Xu, Wang, Xu, and Zhu]{liu-etal-2023-g}
Yang Liu, Dan Iter, Yichong Xu, Shuohang Wang, Ruochen Xu, and Chenguang Zhu.
\newblock {G}-eval: {NLG} evaluation using gpt-4 with better human alignment.
\newblock In Houda Bouamor, Juan Pino, and Kalika Bali, editors,
  \emph{Proceedings of the 2023 Conference on Empirical Methods in Natural
  Language Processing}, pages 2511--2522, Singapore, December 2023. Association
  for Computational Linguistics.
\newblock \doi{10.18653/v1/2023.emnlp-main.153}.
\newblock URL \url{https://aclanthology.org/2023.emnlp-main.153/}.

\bibitem[Lu and Wang(2024)]{Lu2024GenerativeSU}
Xinyi Lu and Xu~Wang.
\newblock Generative students: Using llm-simulated student profiles to support
  question item evaluation.
\newblock \emph{Proceedings of the Eleventh ACM Conference on Learning @
  Scale}, 2024.
\newblock URL \url{https://api.semanticscholar.org/CorpusID:269922100}.

\bibitem[{Maier} et~al.(2025){Maier}, {Aslak}, {Fiaschi}, {Rismal}, {Fletcher},
  {Luhmann}, {Dow}, {Pappas}, and {Wiecki}]{kaiser2025simulating}
Benjamin~F. {Maier}, Ulf {Aslak}, Luca {Fiaschi}, Nina {Rismal}, Kemble
  {Fletcher}, Christian~C. {Luhmann}, Robbie {Dow}, Kli {Pappas}, and Thomas~V.
  {Wiecki}.
\newblock {LLMs Reproduce Human Purchase Intent via Semantic Similarity
  Elicitation of Likert Ratings}.
\newblock \emph{arXiv e-prints}, art. arXiv:2510.08338, October 2025.
\newblock \doi{10.48550/arXiv.2510.08338}.

\bibitem[Martynova et~al.(2025)Martynova, Macina, Daheim, Yalcin, Zhang, and
  Sachan]{martynova-etal-2025-llms}
Daria Martynova, Jakub Macina, Nico Daheim, Nilay Yalcin, Xiaoyu Zhang, and
  Mrinmaya Sachan.
\newblock Can {LLM}s effectively simulate human learners? teachers' insights
  from tutoring {LLM} students.
\newblock In Ekaterina Kochmar, Bashar Alhafni, Marie Bexte, Jill Burstein,
  Andrea Horbach, Ronja Laarmann-Quante, Ana{\"i}s Tack, Victoria Yaneva, and
  Zheng Yuan, editors, \emph{Proceedings of the 20th Workshop on Innovative Use
  of NLP for Building Educational Applications (BEA 2025)}, pages 100--117,
  Vienna, Austria, July 2025. Association for Computational Linguistics.
\newblock ISBN 979-8-89176-270-1.
\newblock \doi{10.18653/v1/2025.bea-1.8}.
\newblock URL \url{https://aclanthology.org/2025.bea-1.8/}.

\bibitem[Noy and Zhang(2023)]{noy2023experimental}
Shakked Noy and Whitney Zhang.
\newblock Experimental evidence on the productivity effects of generative
  artificial intelligence.
\newblock \emph{Science}, 381\penalty0 (6654):\penalty0 187--192, 2023.

\bibitem[{OpenStax, Rice University}(2026)]{openstax2026}
{OpenStax, Rice University}.
\newblock Openstax | free digital textbooks and teaching tools, 2026.
\newblock URL \url{https://openstax.org/}.
\newblock Accessed: 2026-02-14.

\bibitem[Schick et~al.(2023)Schick, Dwivedi-Yu, Dess{\`\i}, Raileanu, Lomeli,
  Hambro, Zettlemoyer, Cancedda, and Scialom]{schick2023toolformer}
Timo Schick, Jane Dwivedi-Yu, Roberto Dess{\`\i}, Roberta Raileanu, Maria
  Lomeli, Eric Hambro, Luke Zettlemoyer, Nicola Cancedda, and Thomas Scialom.
\newblock Toolformer: Language models can teach themselves to use tools.
\newblock \emph{Advances in Neural Information Processing Systems},
  36:\penalty0 68539--68551, 2023.

\bibitem[Schuirmann(1987)]{schuirmann1987comparison}
Donald~J Schuirmann.
\newblock A comparison of the two one-sided tests procedure and the power
  approach for assessing the equivalence of average bioavailability.
\newblock \emph{Journal of pharmacokinetics and biopharmaceutics}, 15\penalty0
  (6):\penalty0 657--680, 1987.

\bibitem[Technavio(2026)]{Technavio2026USprep}
Technavio.
\newblock Test preparation market in us growth analysis - size and forecast
  2025-2029.
\newblock Technical report, Infiniti Research Limited, 2026.
\newblock URL
  \url{https://www.technavio.com/report/test-preparation-market-industry-in-the-us-analysis}.
\newblock Accessed: February 13, 2026.

\bibitem[{Tian} et~al.(2026){Tian}, {Huynh}, {Christhilf}, {Chakraborty},
  {Watanabe}, {Arner}, and {McNamara}]{tian2026cognitively}
Yu~{Tian}, Linh {Huynh}, Katerina {Christhilf}, Shubham {Chakraborty}, Micah
  {Watanabe}, Tracy {Arner}, and Danielle {McNamara}.
\newblock {Cognitively Diverse Multiple-Choice Question Generation: A Hybrid
  Multi-Agent Framework with Large Language Models}.
\newblock \emph{arXiv e-prints}, art. arXiv:2602.03704, February 2026.
\newblock \doi{10.48550/arXiv.2602.03704}.

\bibitem[Wei et~al.(2022)Wei, Wang, Schuurmans, Bosma, Xia, Chi, Le, Zhou,
  et~al.]{wei2022chain}
Jason Wei, Xuezhi Wang, Dale Schuurmans, Maarten Bosma, Fei Xia, Ed~Chi, Quoc~V
  Le, Denny Zhou, et~al.
\newblock Chain-of-thought prompting elicits reasoning in large language
  models.
\newblock \emph{Advances in neural information processing systems},
  35:\penalty0 24824--24837, 2022.

\bibitem[Yao et~al.(2023)Yao, Zhao, Yu, Du, Shafran, Narasimhan, and
  Cao]{yao2022react}
Shunyu Yao, Jeffrey Zhao, Dian Yu, Nan Du, Izhak Shafran, Karthik~R.
  Narasimhan, and Yuan Cao.
\newblock React: Synergizing reasoning and acting in language models.
\newblock In \emph{International Conference on Learning Representations
  (ICLR)}, 2023.
\newblock URL \url{https://openreview.net/forum?id=WE_vluYUL-X}.

\bibitem[Zhang et~al.(2024)Zhang, Chen, Jin, Wang, Ji, Wang, and
  Han]{zhang-etal-2024-comprehensive-survey}
Yu~Zhang, Xiusi Chen, Bowen Jin, Sheng Wang, Shuiwang Ji, Wei Wang, and Jiawei
  Han.
\newblock A comprehensive survey of scientific large language models and their
  applications in scientific discovery.
\newblock In Yaser Al-Onaizan, Mohit Bansal, and Yun-Nung Chen, editors,
  \emph{Proceedings of the 2024 Conference on Empirical Methods in Natural
  Language Processing}, pages 8783--8817, Miami, Florida, USA, November 2024.
  Association for Computational Linguistics.
\newblock \doi{10.18653/v1/2024.emnlp-main.498}.
\newblock URL \url{https://aclanthology.org/2024.emnlp-main.498/}.

\bibitem[Zheng et~al.(2023)Zheng, Chiang, Sheng, Zhuang, Wu, Zhuang, Lin, Li,
  Li, Xing, et~al.]{zheng2023judging}
Lianmin Zheng, Wei-Lin Chiang, Ying Sheng, Siyuan Zhuang, Zhanghao Wu, Yonghao
  Zhuang, Zi~Lin, Zhuohan Li, Dacheng Li, Eric Xing, et~al.
\newblock Judging llm-as-a-judge with mt-bench and chatbot arena.
\newblock \emph{Advances in neural information processing systems},
  36:\penalty0 46595--46623, 2023.

\bibitem[Zheng et~al.(2025)Zheng, Deng, Tsang, Wang, Bai, Wang, and
  Song]{zheng2025automation}
Tianshi Zheng, Zheye Deng, Hong~Ting Tsang, Weiqi Wang, Jiaxin Bai, Zihao Wang,
  and Yangqiu Song.
\newblock From automation to autonomy: A survey on large language models in
  scientific discovery.
\newblock \emph{arXiv preprint arXiv:2505.13259}, 2025.

\end{thebibliography}

%\newpage
\appendix
\appendixsectionformat
\section{AI Tools}
\label{app:ai-tools}

This study was conducted using a coordinated multi-agent workflow that integrated 
Claude Code\footnote{\url{https://claude.ai/code}} (Anthropic),
Codex\footnote{\url{https://openai.com/codex}} (OpenAI), 
Gemini CLI\footnote{\url{https://github.com/google-gemini/gemini-cli}} (Google), 
and Ampcode\footnote{\url{https://ampcode.com/}} (Amp) 
as development and research assistants.

The underlying model and tooling stack included:
\begin{enumerate}
    \item \textbf{Google AI Studio Gemini 3 Pro Preview}, used for SAT question extraction from
  PDF source files.
    \item \textbf{OpenAI GPT-5-nano} and \textbf{Google Gemini 2.5 Flash}, used for MCQ
  generation and LLM-based evaluation.
    \item \textbf{Mistral OCR 3}, used to convert textbook PDFs into Markdown for downstream
  processing.
\end{enumerate}

\section{Time and Cost}
\label{app:cost}

The project was completed over 10 days, from February 2, 2026 to February 12, 2026, by a single
researcher who orchestrated agents and managed all LLM interactions. On each day (including
weekends), the researcher spent approximately 8 hours on project activities.

The tooling and billing setup was as follows:
\begin{itemize}
    \item Anthropic Claude Code subscription: \$20/month
    \item OpenAI Codex subscription: \$20/month
    \item Google Gemini API: pay-as-you-go
    \item OpenAI API: pay-as-you-go
    \item Mistral API: pay-as-you-go
    \item Ampcode: \$10 daily free with ads 
\end{itemize}

Direct usage costs incurred during this project were:
\begin{enumerate}
      \item \$0 for extracting SAT Math questions from PDFs downloaded from \url{https://
  satsuitequestionbank.collegeboard.org/}.
      \item \$11 for converting four textbook PDFs to Markdown via Mistral OCR.
      \item \$9 for OpenAI API usage.
      \item \$42 for Gemini API usage.
\end{enumerate}

\section{Datasets}
\label{app:datasets}

Table \ref{tab:datasets} describes the three sets of data in the study.

\begin{table}[h]
\centering
\caption{Datasets}
\label{tab:datasets}
\small
\begin{tabularx}{\textwidth}{@{}l X X c@{}}
\toprule
Artifact & Description & Source & N \\
\midrule
SAT baseline 
& Expert-vetted SAT Math MCQs (text extracted from PDFs; incomplete items filtered) 
& College Board SAT Suite Question Bank PDFs 
& $\sim$1{,}017 \\

Gemini-generated 
& MCQs generated from open textbook chunks under domain, skill, difficulty, and cognitive-level constraints 
& LLM-based generator (Gemini) 
& $\sim$1{,}065 \\

GPT-generated 
& MCQs generated from open textbook chunks under the same constraints 
& LLM-based generator (GPT) 
& $\sim$1{,}071 \\
\bottomrule
\end{tabularx}
\end{table}

\section{Evaluation rubric (24 criteria)}
\label{app:rubric}

Table~\ref{tab:evaluation-criteria} lists
the evaluation criteria. 
Each criterion is scored on a 1--5 scale by an LLM judge.

\begin{itemize}
\item \textbf{5} = Excellent / no issues
\item \textbf{4} = Minor issues, still strong
\item \textbf{3} = Noticeable issues that could affect learning/validity
\item \textbf{2} = Serious issues; should be revised
\item \textbf{1} = Fundamentally broken / invalid
\end{itemize}

\noindent Definitions below are paraphrased for readability.

\begin{longtable}{@{}p{0.30\textwidth}p{0.22\textwidth}p{0.44\textwidth}@{}}
\toprule
\textbf{Criterion} & \textbf{Group} & \textbf{Short definition (paraphrased)} \\
\midrule
\endhead
\texttt{factual\_accuracy} & Accuracy \& Grounding & Stem/options/key are mathematically correct as written. \\
\texttt{no\_hallucinations} & Accuracy \& Grounding & Does not invent missing context; no unsupported claims. \\
\texttt{correct\_key\_answer} & Accuracy \& Grounding & The labeled correct option is truly correct. \\
\texttt{content\_alignment} & Accuracy \& Grounding & Item targets the intended concept from the mapped chunk. \\
\midrule
\texttt{clarity\_question} & Clarity \& Presentation & Stem is complete, readable, and unambiguous. \\
\texttt{clarity\_options} & Clarity \& Presentation & Options are parallel, readable, and distinguishable. \\
\texttt{grammar\_fluency} & Clarity \& Presentation & Language and notation are fluent and consistent. \\
\texttt{no\_duplicates} & Clarity \& Presentation & No duplicate or near-duplicate options. \\
\midrule
\texttt{no\_answer\_hints} & Distractor \& Answer Quality & No inadvertent clues revealing the answer. \\
\texttt{plausible\_distractors} & Distractor \& Answer Quality & Wrong options are plausible and misconception-based. \\
\texttt{unambiguous\_correctness} & Distractor \& Answer Quality & Exactly one defensible correct answer. \\
\texttt{recall\_vs\_guessing} & Distractor \& Answer Quality & Requires reasoning/knowledge, not superficial guessing. \\
\midrule
\texttt{difficulty\_balance} & Difficulty \& Engagement  & Difficulty is appropriate and not trick-inflated/deflated. \\
\texttt{cognitive\_engagement} & Difficulty \& Engagement  & Requires meaningful cognitive work beyond recognition. \\
\texttt{holistic\_quality} & Difficulty \& Engagement & Overall quality combining correctness, clarity, and pedagogy. \\
\midrule
\texttt{domain\_alignment} & Domain \& Skill Alignment & Declared SAT domain matches actual content. \\
\texttt{domain\_appropriateness} & Domain \& Skill Alignment & Domain label is reasonable for the problem. \\
\texttt{skill\_alignment} & Domain \& Skill Alignment & Declared skill matches what is tested. \\
\texttt{skill\_depth} & Domain \& Skill Alignment & Measures depth (conceptual understanding, not pattern match). \\
\texttt{skill\_isolation} & Domain \& Skill Alignment & Primarily tests the declared skill without confounds. \\
\midrule
\texttt{difficulty\_alignment} & Difficulty \& Cognitive Calibration  & Matches the declared difficulty label. \\
\texttt{difficulty\_consistency} & Difficulty \& Cognitive Calibration  & No hidden leaps; consistent difficulty across stem/options. \\
\texttt{cognitive\_level\_alignment} & Difficulty \& Cognitive Calibration & Bloom level label matches the item’s demand. \\
\texttt{cognitive\_level\_ appropriateness} & Difficulty \& Cognitive Calibration & Bloom level is appropriate for SAT context and intent. \\
\bottomrule
\caption{Evaluation Criteria}
\label{tab:evaluation-criteria}
\end{longtable}

\section{Statistical analysis}
\label{app:stats}

\subsection{Two comparison tracks}
\begin{enumerate}
\item \textbf{Matched track (paired):} Compare only shared \texttt{question\_id}s across sets.
\item \textbf{Whole-set track (unpaired):} Compare all available MCQs per set.
\end{enumerate}

\subsection{Primary inferential tests (summary)}
\begin{itemize}
\item Matched: Friedman omnibus; Wilcoxon signed-rank pairwise (Holm correction); paired effect size ($d_z$); paired TOST equivalence.
\item Whole-set: Welch ANOVA (or Kruskal--Wallis fallback); Welch pairwise (Holm correction); Hedges $g$; independent-sample TOST equivalence.
\end{itemize}

\subsection{Similarity decision rules}
\begin{itemize}
\item \textbf{Strict similarity:} 24/24 criteria equivalent.
\item \textbf{Practical similarity:} $\ge$19/24 criteria equivalent.
\end{itemize}

\section{Research Activity Logs}
\label{app:activity_logs}

A coarse-grained log file documenting the daily research activities is publicly available 
in the project repository at: \url{https://github.com/anyuanay/context_aware_sat_math_mcqs/blob/main/docs/plans/Research_Activity_Log.md}.

\section{Reproducibility}
\label{app:reproducibility}

To support full reproducibility, all project artifacts, including prompts, datasets, and analysis
code, are publicly available at: \url{https://github.com/anyuanay/context_aware_sat_math_mcqs}.

\end{document}